\documentclass[pdflatex,sn-mathphys]{sn-jnl}

\jyear{2022}%

\theoremstyle{thmstyleone}%
\theoremstyle{thmstyletwo}%

\theoremstyle{thmstylethree}%

\begin{document}

\title[Channeling: a new class of dissolution in complex porous media]{Channeling: a new class of dissolution in complex porous media}


\author*[1]{\fnm{Hannah P.} \sur{Menke}}\email{h.menke@hw.ac.uk}
\equalcont{These authors contributed equally to this work.}

\author[1]{\fnm{Julien} \sur{Maes}}\email{j.maes@hw.ac.uk}
\equalcont{These authors contributed equally to this work.}

\author[2]{\fnm{Sebastian} \sur{Geiger}}\email{s.geiger@tudelft.nl}

\affil*[1]{\orgdiv{Institute of GeoEnergy Engineering}, \orgname{Heriot-Watt University}, \orgaddress{ \city{Edinburgh}, \country{United Kingdom}}}

\affil[2]{\orgdiv{Department of Geoscience and Engineering}, \orgname{Delft University of Technology}, \orgaddress{ \country{Netherlands}}}


\abstract{The current conceptual model of mineral dissolution in porous media is comprised of three dissolution patterns (wormhole, compact, and uniform) - or regimes - that develop depending on the relative dominance of flow, diffusion, and reaction rate. Here, we examine the evolution of pore structure during acid injection using numerical simulations on two porous media structures of increasing complexity. We examine the boundaries between regimes and characterise the existence of a fourth regime called channeling, where already existing fast flow pathways are preferentially widened by dissolution. Channeling occurs in cases where the distribution in pore throat size results in orders of magnitude differences in flow rate for different flow pathways. This focusing of dissolution along only dominant flow paths induces an immediate, large change in permeability with a comparatively small change in porosity, resulting in a porosity-permeability relationship unlike any that has been previously seen. This work demonstrates that our current conceptual model of dissolution regimes must be modified to include channeling for accurate predictions of dissolution in applications such as geologic carbon storage and geothermal energy production. }




\maketitle

\section{Introduction}\label{Intro}

The current conceptual model of mineral dissolution in porous media is based on three 'dissolution regimes' that assist flow and transport prediction during dissolution\cite{1998-Fredd,2002-Golfier,1986-Chadam}. Accurate identification of these regimes is essential as the dissolution regime ultimately controls the evolution of permeability. Moving from one regime to the other results in orders of magnitude differences in permeability change with increasing porosity. As such, accurate prediction of mineral dissolution in porous media is crucial for a wide range of subsurface applications, including CO$_2$ sequestration and geothermal power generation \cite{2015-Pandey,2015-Black} where failure to predict the changes in permeability can lead to poor fluid injection efficiency and potentially irreversible reservoir damage \cite{2010-Gauss,2010-Portier}.

\begin{figure}[!t]
\includegraphics[width=1\textwidth]{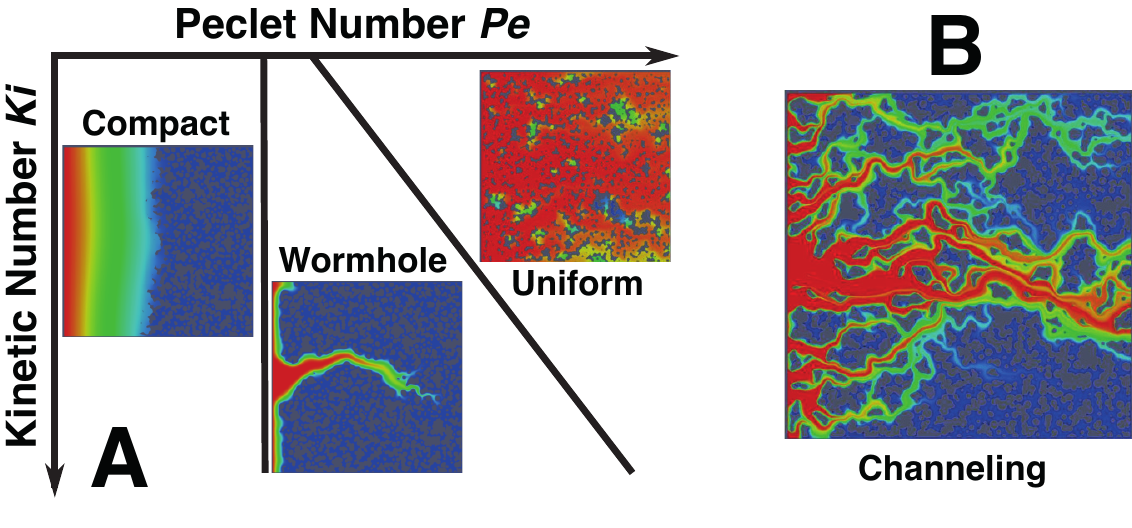}
\caption{(A) Schematic depiction of dissolution regimes in the P\'eclet number - Kinetic number space. (B) This paper modifies this traditional conceptual model by adding the channeling regime. \label{fig:conceptualmodel}}
\end{figure}

The balance between flow, diffusion, and reaction rates determines which dissolution pattern develops during reactive flow in a porous medium \cite{2013-szymczak}. When flow is slow compared to reaction rate, the face of the porous medium closest to the inlet will dissolve and result in compact dissolution. When flow is fast compared to the reaction rate, acidic fluid is quickly distributed throughout the pore spaces and the medium dissolves uniformly. At intermediate flow rates, the acidic fluid etches a wide pathway through the porous medium in the direction of flow and forms a wormhole. These regimes can be predicted based on the P\'eclet number $Pe$ (the ratio of advective to diffusive transport) and the Kinetic number $Ki$ (the ratio of chemical reaction to diffusive transport). However, these dissolution regimes do not take into account the structural heterogeneity of complex porous media, because they were first identified (Fig \ref{fig:conceptualmodel}) before the technology was developed to observe or model reactive flow at the scale of grains and pores. Thus, they are problematic when quantifying the relationships between flow, reaction, and pore structure.

Recent advances in x-ray-CT imaging techniques \cite{2017-reynolds, menke2022using} have enabled direct observation and quantification of dissolution-induced changes in the pore structure and provided insight into influences of structural heterogeneity, flow, and reaction rate on dissolution regime. Several experimental studies have observed mineral dissolution at the pore-scale in reservoir rock samples \cite{2009-Noirel,2013-Hao,2014-Luquot, 2015-deng, garing2015anti}. Others \cite{2015-Menke,2016b-Menke,2017-Menke,2018-Menke} studied the dissolution dynamics \textit{in situ} during fast flow in rocks of varying complexity, observing uniform dissolution in a structurally simple rock, but the opening of preferential flow pathways in the more complex rock samples. This path-widening did not progress longitudinally with flow, as is the case for wormholes, but instead opened everywhere along the dominant flow channel and was thus named 'channeling'. This regime was later confirmed \cite{2020-Yang} by observations of channeling in both fractured and vuggy rock samples. However, as of yet no in-depth experimental characterisation of the conditions required for channeling has been performed, and thus no new conceptual model has been proposed that includes channeling. 

Pore-scale experimental techniques are often complemented by advances in numerical simulations that give insight into the complex relationship between pore structure, flow, and reaction. However, limitations in the numerical methods have not allowed for flow to be simulated at the high flow rates seen near reservoir injection wells \cite{2016-nunes,2016b-nunes,2016-Gray}, which limits the range of dissolution regimes that can be studied. Several studies \cite{2009-Szymczak, 2017-Soulaine} have attempted a comprehensive numerical investigation of the full spectrum of pore-scale dissolution regimes (Fig. \ref{fig:conceptualmodel}), but these were restricted to relatively homogeneous domains with minor differences in pore structure between models and small differences in flow rate. Channeling has thus not been characterised in numerical models at the pore-scale by any study to date because either the numerical capabilities for high flow rates or structural complexity in the model were lacking. Therefore, the placement of the boundaries between wormhole, channeling, and uniform dissolution regimes are unknown and the conceptual model of dissolution is missing information vital for accurate modelling of dissolution.

The work presented here is a numerical investigation into how pore-space complexity changes the conceptual model of dissolution regimes and how the channeling regime fits into our broader understanding of dissolution. Two synthetic 2D pore structures with varying levels of heterogeneity were created stochastically and their structural complexity characterized (Fig. \ref{fig:micromodels}). A series of 26 numerical simulations was performed on each of the geometries by injecting acid at different flow and reactive conditions using our new highly efficient open source numerical solver GeoChemFoam \cite{2021a-Maes,2021b-Maes,2021c-Maes,2022a-Maes}, which is based on the Open Source Computational Fluid Dynamics toolbox OpenFOAM \cite{2016-OpenFOAM}. We observe that many of the model scenario results do not fit the conceptual model of the three traditional dissolution regimes and have fundamentally categorically distinct porosity-permeability relationships.  We show that these four dissolution regimes can be distinguished using the moments (mean, standard deviation, skewness, and kurtosis) of the distributions of pore throat size and acid concentration. We then employ hierarchical agglomerative clustering \cite{1987-ROUSSEEUW} to provide a quantitative measure of identifying the channeling regime and differentiating channeling from the other three regimes. Finally, we provide an updated conceptual model of dissolution regimes that includes channeling and demonstrate how the boundaries between regimes shift with increasing pore space complexity. 

\section{Results}

\subsection{Numerical observations of pore-scale dissolution}\label{Sect:images}

\begin{figure*}
\begin{center}
\includegraphics[width=0.95\textwidth]{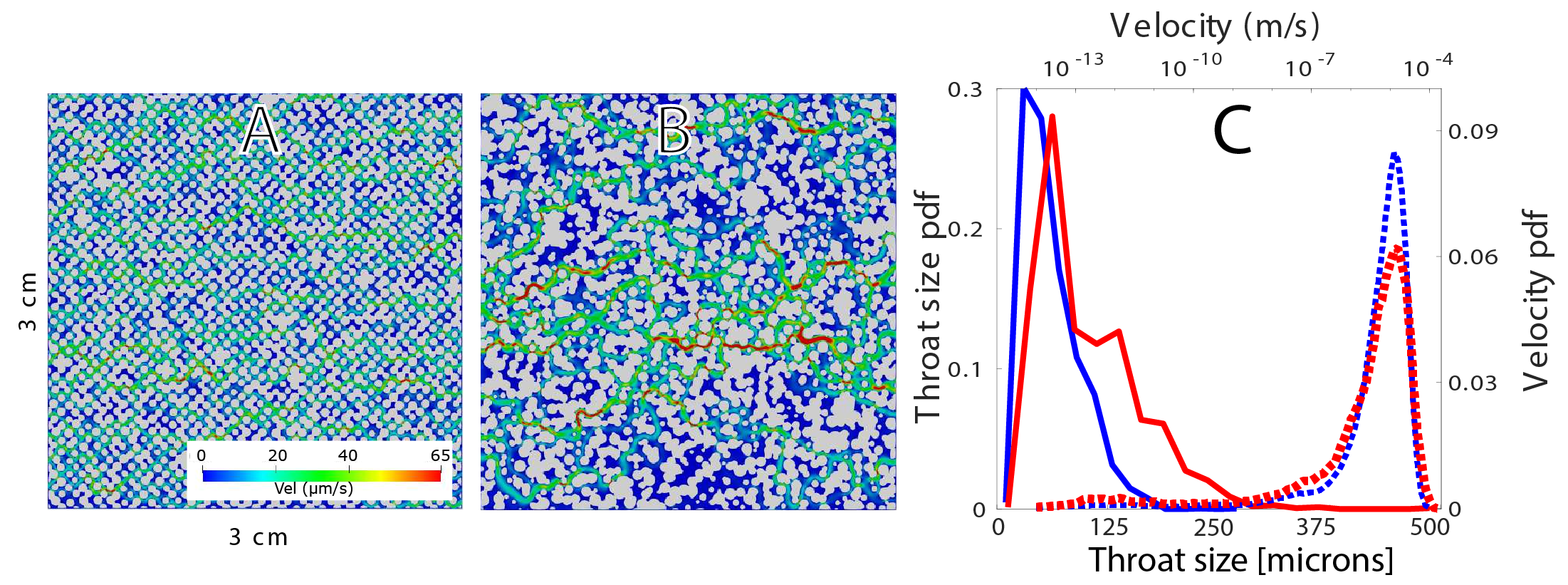}
\caption{(A) Model A. (B) Model B. The grains (gray) are rendered with the velocity field of the pore space (color) computed using an injection rate of 0.4 mL/min and a resolution of 2.5 $\mu$m per pixel. (C) Results showing the histogram of throat size (solid) and velocity (dotted) for Model A (blue) and Model B (red). The characteristic length $L$ is 1.125 x $10^{-4}$ m for Model A and 1.251 x $10^{-4}$ m for Model B. \label{fig:micromodels}}
\end{center}
\end{figure*}

A relatively homogeneous geometry was created with a small random deviation in both grain radius and placement of the grains (Model A, Fig \ref{fig:micromodels}A). Structural complexity was then increased by adding a larger random deviation of both grain radius and placement to create an increasingly heterogeneous geometry (Model B, Fig \ref{fig:micromodels}B). The distributions of throat sizes and velocity of Model A and Model B are presented in Fig \ref{fig:micromodels}C. Model A has velocity and pore throat size distributions that are narrow, while Model B shows a wide tail representing the focusing of flow into the preferential flow paths through larger pore throats. Additional details on geometry creation and the numerical modelling are included in the supplementary material.

For each geometry, we perform 26 simulations to identify the boundaries between dissolution regimes. The model solves the quasi-steady state Navier-Stokes equations and advection-diffusion of reactant in the pore space using a finite-volume discretization on an unstructured hybrid mesh consisting of hexahedral and split-hexahedral elements \citep{2016-OpenFOAM}. The numerical model, including meshing, time-stepping and convergence, is presented in detail in the supplementary material. A simplified chemical model is employed representing dissolution of calcite mineral during acid injection, with one fluid component and one reaction component \cite{2016-nunes,2017-Soulaine,2022a-Maes}. The molecular diffusion is the constant $D=10^{-9}$ m$^2$.s$^{-1}$. The displacement of the fluid-solid interface is handled using the Arbitrary Eulerian Lagrangian (ALE) method. Acid is injected from the left boundary at constant concentration and flow rate and the simulations are ended either when the porosity increases to 1.6 times the initial porosity or the permeability reaches a value 100 times larger than the initial permeability.  

The relative importance of advection and reaction rate to molecular diffusion is characterized by the P\'eclet number $Pe=UL/D$ and Kinetic number $Ki=kL/D$, where $U$ [m.s$^{-1}$] is the average pore velocity, $L$ [m] is the average width of the flow pathways and $k$ [m.s$^{-1}$] is the reaction constant. Details on how to calculate $Pe$, $Ki$, $U$ and $L$ are presented in the supplementary material. For each simulation, the flow rate and reaction constant are adjusted to obtain the desired $Pe$ and $Ki$ at time=0.

\begin{figure*}[!t]
\begin{center}
\includegraphics[width=1\textwidth]{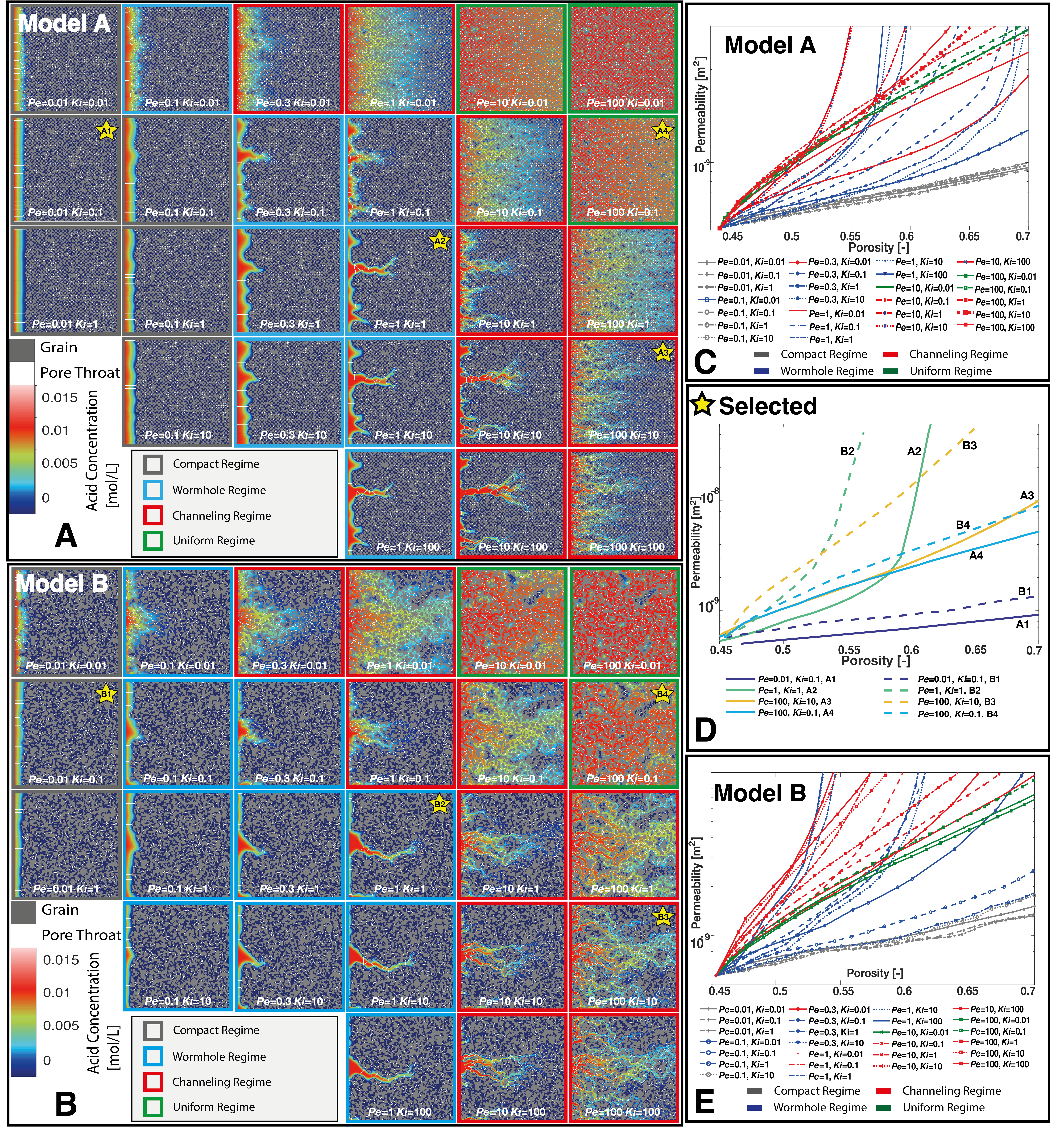}
\caption{Pore structure and acid concentration during mineral dissolution in (A) Model A and (B) Model B at $Pe$ and $Ki$ ranging from 0.01 to 100 at a porosity of 0.57. The solid phase is rendered in grey and the acid concentration in colors. The pore throats extracted using a watershed algorithm are shown in white. Simulations categorised in the compact, wormhole, and uniform regimes are outlined in gray, blue, and green, respectively, while simulations that do not fit into any traditional regime are outlined in red and designated channeling. (C,E) Porosity-Permeability curves for all the Model A and Model B simulations, respectively. (D) Porosity-Permeability curves for selected (starred) simulations. \label{fig:GeometryA}}
\end{center}
\end{figure*}

Maps showing the distribution of the injected acid concentration at the time where dissolution has increased the porosity from ~0.45 to ~0.5 are presented (Fig. \ref{fig:GeometryA}A and B).  Videos of the dynamic evolution of dissolution are provided in the supplementary material. In Fig \ref{fig:GeometryA}A, we observe the three traditional regimes for Model A: compact dissolution (gray), wormhole (blue) and uniform dissolution (green). The cases at the boundary between wormhole and uniform dissolution, outlined in red, are traditionally classified as (ramified) wormholes  \cite{2009-Szymczak,2017-Soulaine}. However, here we observe they exhibit characteristics that contradict the wormholing concept. Rather than one ramified wormhole that has very little change in permeability until breakthrough (e.g. $Pe=1, Ki=0.1$), these include a very large number of small dissolution channels that extend towards the outlet of the model, resulting in a porosity-permeability evolution with similar curvature to those of uniform dissolution, but with a larger change in permeability with porosity as dissolution is present in these pathways at the outlet almost instantaneously. In these cases, there is a direct correspondence between dissolution pathways and initial fast flow paths (Fig. \ref{fig:micromodels}A). The most dominant flow paths are dissolved first, which leads to an initial increase in permeability that is higher than that observed for uniform dissolution (e.g. $Pe=100, Ki=10$) (Fig. \ref{fig:GeometryA}A). We will demonstrate that this regime is channeling, as identified in previous experimental studies \cite{2016b-Menke,2020-Yang}. 

\begin{figure*}[!t]
\begin{center}
\includegraphics[width=0.9\textwidth]{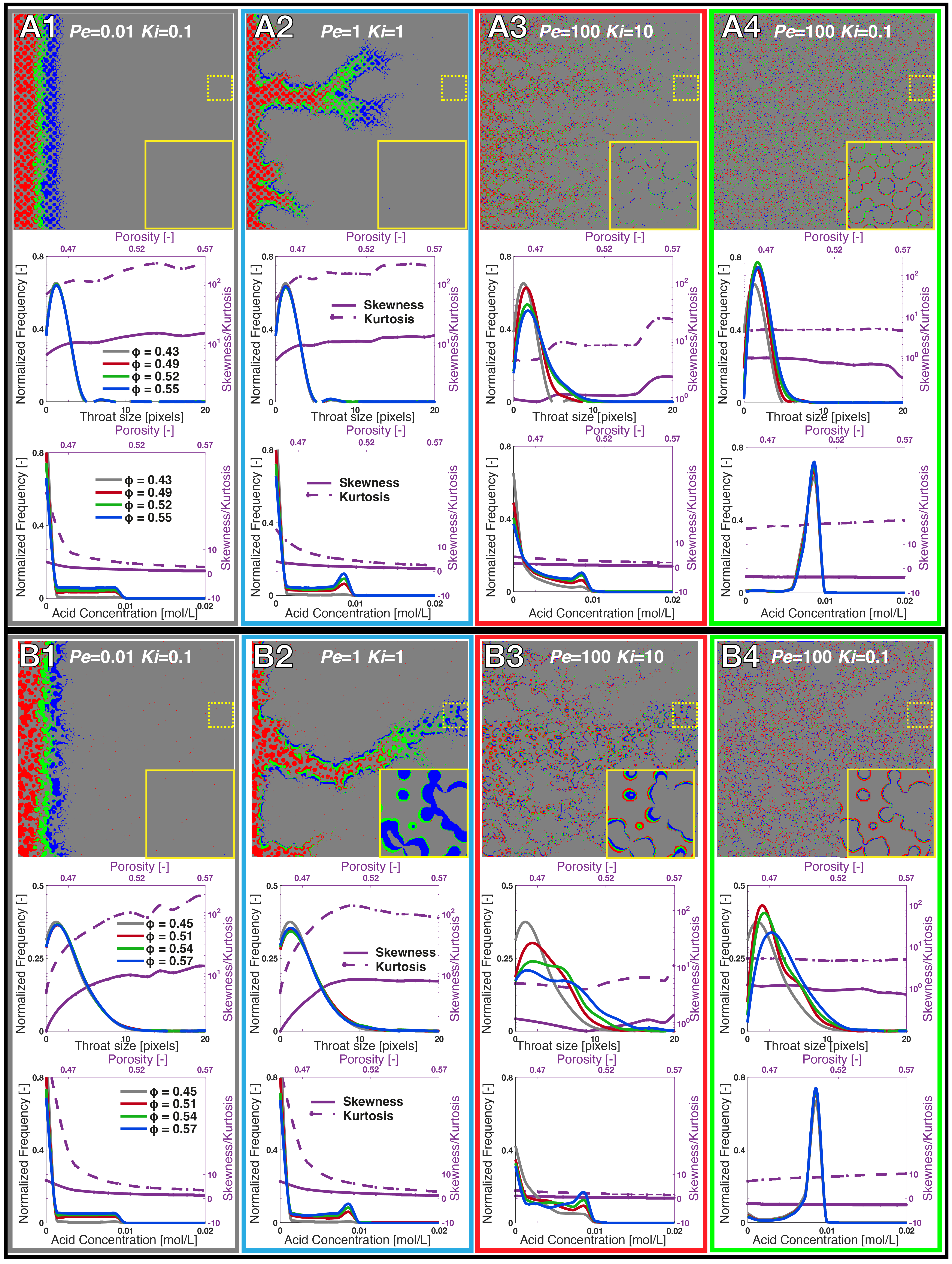}
\caption{(Rows 1 and 4) Dissolution with change in porosity as a proxy for time for select simulations of Model A (A1-4) and B (B1-4). In the top of each example, the dissolved pore space is shown at different porosity values with undissolved being grey, and red, green, and blue showing dissolution at subsequent times. The solid yellow squares are magnified regions of interest of the outlet of the model during dissolution which are outlined as dashed yellow squares. (Rows 2 and 4) The pore throat size distributions of Model A (row 2) and Model B (row 4) of the above simulation at the porosity values depicted in rows 1 and 4 (grey, red, green, blue). The skewness (solid line) and kurtosis (dashed line) of pore throat size are shown with increasing porosity on the right axes (purple). (Rows 3 and 6) The concentration distributions of Model A (row 3) and Model B (row 4) at the same porosity values depicted in rows 1 and 4 (grey, red, green, blue). The skewness (solid line) and kurtosis (dashed line) of the concentration distributions are shown with increasing porosity using the right axes (purple). \label{fig:DissolutionThroatSize}}
\end{center}
\end{figure*}

The existence of channeling becomes more apparent as structural complexity increases in Model B (Fig. \ref{fig:GeometryA}B), where we again observe a number of cases (outlined in red) that cannot be classified using any of the three traditional regimes and instead follow the same convex porosity-permeability (Fig. \ref{fig:DissolutionThroatSize}E) trends as those in Model A (Fig. \ref{fig:DissolutionThroatSize}C). In addition, the increased structural complexity has increased the order of the porosity-permeability change with faster widening of the channels in the more heterogeneous cases. In all of the channeling cases, the permeability increases faster and attains a larger value than for uniform dissolution and is faster than for the more structurally homogeneous cases in Fig. \ref{fig:GeometryA}A.

To illustrate the impact of pore space heterogeneity on dissolution regime, the evolution of the dissolution patterns and the throat size and concentration distributions for selected cases of Model A and B are shown (Fig. \ref{fig:DissolutionThroatSize}D). The details of image analysis techniques used to extract these metrics can be found in the supplementary material. The corresponding evolutions of the porosity-permeability relationships are shown in Fig. \ref{fig:GeometryA}D.  In the compact dissolution cases (A1, B1), the dissolution is transport-limited and creates large throats at the front of the model that result in a large skewness and kurtosis in throat size. Conversely, as the dissolution front advances, the highly concentrated acid spread into more of the pore space and the skewness and kurtosis of the concentration distributions decrease.  The small deviations in the dissolution front in Model B result in a larger overall skewness and kurtosis of throat size and concentration and a larger slope in the porosity-permeability relationship than Model A. However, even at the largest porosity shown for Model B (porosity = 0.57), the dissolution front remains stable, and there is no dissolution near the outlet, so the overall change in permeability is low. When we apply a power law fit to the porosity-permeability relationship, we find the relative small exponent of 1.5 to 2. 

In the cases A2 and B2, the dissolution front becomes unstable, and advection and reaction compete as the dissolution etches pathways (wormholes) through the models. Large pore throats are created both at the fronts and inside the wormholes, which result in a large increase in the skewness and kurtosis of pore throat size and a widening of the pore throat size distribution through time. The concentration distributions develop a peak indicative of a preferential flow path through the model with corresponding decreases in skewness and kurtosis, as the wormhole carries the acid towards the outlet. In Model A, similar competition between flow paths results in a porosity-permeability relationship (Fig. \ref{fig:GeometryA}D) that is similar to compact dissolution. There is a rapid increase in permeability once the wormhole is established in the fastest flow pathway, but has a much higher exponent of 11. The preferential flow path is more dominant in Model B. We observe less competition initially with breakthrough of the wormhole to the outlet occurring earlier with a larger increase in permeability and an exponent as high as 19.

In cases A3 and B3 the pore throats in the preferential flow pathways are dissolved across the entire domain at the very beginning of the simulations, which creates a fat tail in the throat size distributions and peaks in the concentration distributions. Notably, the skewness and kurtosis of concentration show very little change due to the broad spread of the acid even from the beginning of the simulations. Flow is focused in these channels and there is little dissolution in the slower flowing areas of the pore space. This focused dissolution results in a porosity-permeability relationship of power law exponent 6 to 12.  In Model B the structural complexity is higher, and there are fewer fast flowing channels, however, they are more important and result in a higher order porosity-permeability relationship. The dissolution converges towards these fast channels and the flow inside them becomes so dominant that no wormhole forms in the domain. For channeling, flow is stable and the dissolution channels are instantaneously established as the dominant flow pathways and then become wider as the porosity increases. 

In cases A4 and B4 the dissolution is reaction-limited and uniform across the domains, with no preferential pathways forming in either Model A or B. The kurtosis and skewness of pore throat size across the domains is flat as all flow paths are widened together. The concentration distribution has a large peak at the injection concentration which increases only slightly throughout the simulations as more of the model dissolves. Here, the skewness of concentration is below 0, which is contrary to all other dissolution regimes. During uniform dissolution the increased structural heterogeneity results in only a small increase in the power law exponent of the porosity-permeability relationship from 5 to 6. 

\subsection{Channeling: a new class of dissolution regime}\label{Channeling}
\begin{figure}
\includegraphics[width=1\textwidth]{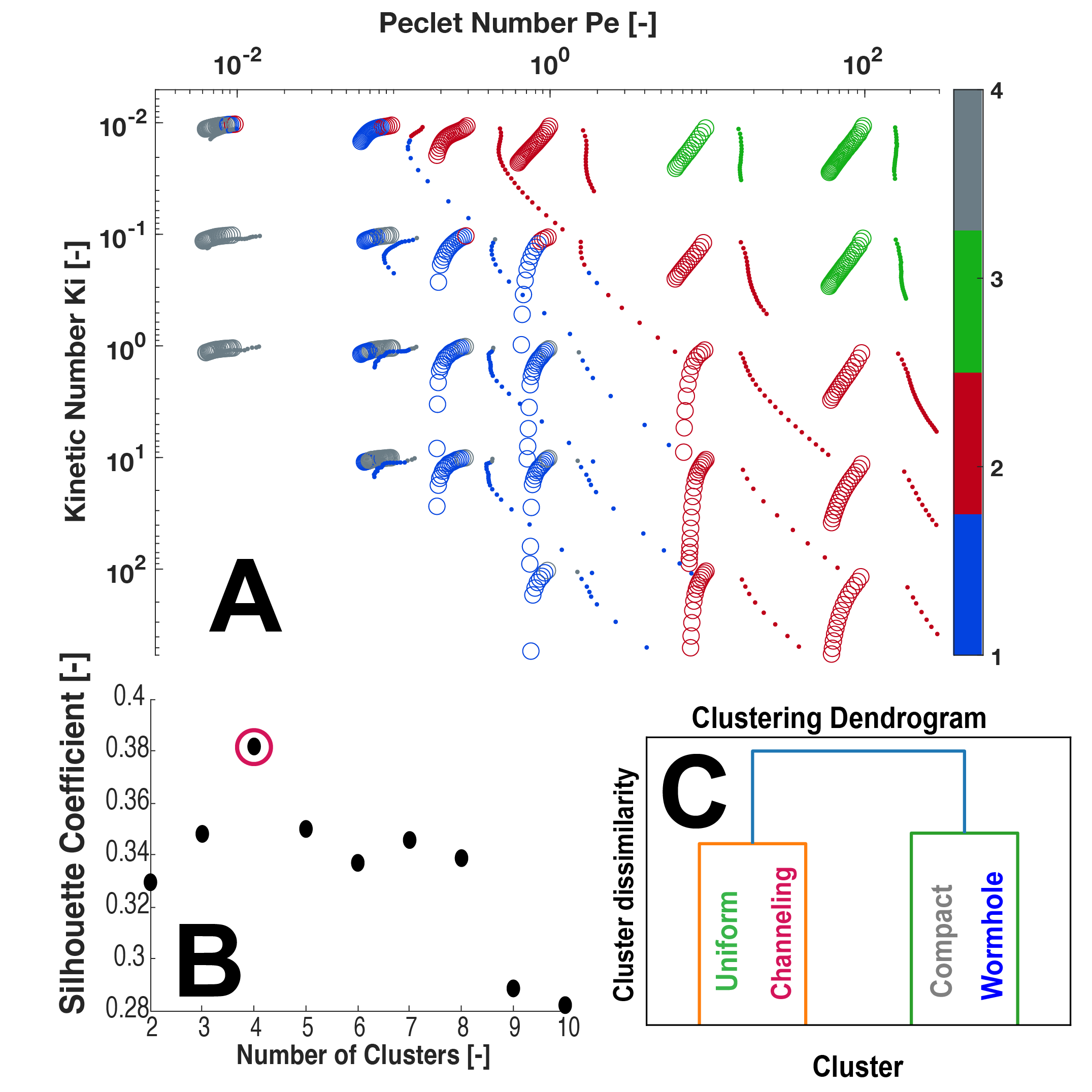}
\caption{Hierarchical Agglomerate Clustering of dissolution using the mean, standard deviation, skewness, and kurtosis of concentration and pore throat size. (A) Each circle (Model A) and dot (Model B) represents a single time-step in one numerical experiment, and is color coded by cluster by the clustering algorithm, according to the dendrogram in (C). (B) Silhouette Coefficient of the number of clusters. (C) The dendrogram of the cluster labels.    \label{fig:Clustering}}
\end{figure}

We quantitatively identify dissolution regime by clustering the four moments (mean, standard deviate, skewness, and kurtosis) of the distributions in concentration and throat size at each time step (Fig \ref{fig:Clustering}).  We used hierarchical agglomerative clustering for a range of numbers of clusters from 2 to 10 shown in Fig \ref{fig:Clustering}B. The Silhouette Coefficient (SC) is used to rank the optimal number of clusters where a higher index indicates that clusters are dense and well separated. For our group of simulations, the highest SC was observed with 4 clusters. This clustering (Fig. \ref{fig:Clustering}A) identifies channeling as independent regime and is in agreement with our visual characterization and physical understanding of the numerical experiments.  

The clustering dendrogram (Fig. \ref{fig:Clustering}C) gives insight into how the clustering algorithm determines each cluster boundary. First the channeling/uniform regimes split from the wormhole/compact regimes, followed by the compact and wormhole regimes, and finally channeling and uniform regimes. The order of splitting indicates that the difference in dissolution behavior is greatest between the channelling/uniform regimes and the wormhole/compact regimes, and smallest between the channeling and uniform regimes, which confirms our assertion that channeling is distinct from wormhole formation. The clustering also indicates that some simulations  straddle the boundary between regimes, beginning in one regime and ending in another as the dissolution changes the distribution of flow within the porous medium and flow becomes more or less stable in preferential flow pathways. This is consistent with our analysis of the dissolution progress shown in Figs \ref{fig:GeometryA} and \ref{fig:DissolutionThroatSize}. 

\begin{figure}
\includegraphics[width=1\textwidth]{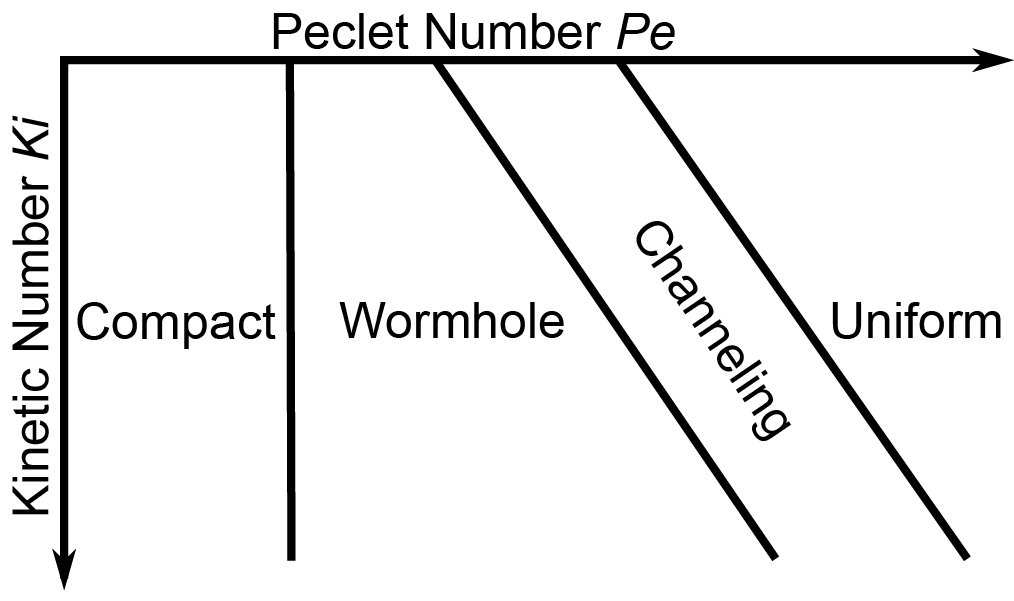}
\caption{The P\'eclet number - Kinetic number space with updated dissolution regimes. Channeling exists as a distinct regime between wormhole and uniform. \label{fig:KiPeWormholeChannel}}
\end{figure}

We present our updated conceptual model of dissolution regimes in Fig \ref{fig:KiPeWormholeChannel}. Channeling is a distinct regime between wormhole formation and uniform dissolution. In more heterogeneous structures, the relative importance of already existing flow paths increases, leading to the formation of wormholes and channels faster, with a higher order. 

\section{Discussion: Reconciling the pore scale with the continuum scale}

We have characterised the dissolution regime of channeling, identified its location within the $Pe-Ki$ space, and quantified its relationship to wormhole formation and uniform dissolution. Previous experimental work in 3D has reported the porosity-permeability of channeling to have a power law order of between 7 and 11 \cite{2016-Menke} and the uniform regime to have an order of 5 \cite{2015-Menke}, which is consistent with our 2D observations of power law order 6 to 12 for channeling and 5 to 6 for the uniform regime. This indicates that the 2D results are likely to be directly extendable to 3D. 

Characterisation of dissolution regimes are crucial for providing accurate porosity-permeability relationships for Darcy and reservoir-scale models. In contrast to other pattern formations such as viscous fingering in multi-phase flow, both the location and the conditions under which dissolution follows pre-existing flow paths is important. Wormholes develop from the pore-scale as micron-scale ramifications that merge and expand to eventually form dissolution pathways that impact flow at the field-scale. Similarly channeling will influence flow during dissolution from the pore- to the field-scale provided that scale-dependent structural complexity exists, as for example with the presence of vugs, fractures and faults \cite{2020-Yang}. Predicting such development of dissolution patterns at the field-scale requires an accurate estimation of the evolving permeability of the dissolving matrix \cite{2020-Faris}. 

This unique research provides a first-ever characterisation of the channeling regime. Channeling occurs in heterogeneous porous media, where differences in pore throat sizes cause dissolution to widen preferential flow pathways. This study is the first step towards understanding the multi-scale interactions between structure and dissolution in more complex multi-scale domains such as carbonate rocks where knowledge of how the pore space dissolves at the scale of grains and pores can be incorporated into field scale models. Indeed, in the carbonate reservoirs typically considered for industrial geologic carbon storage applications with a representative calcite reaction constant and carbonate reference pore throat sizes \cite{2017-Menke}, $Ki$ will range between 0.1 and 100. Therefore at sufficiently fast flow rates, the dissolution will be in the channeling regime. Accurate characterisation of the channeling regime is thus vital for accurate prediction of dissolution during many commercial processes essential for the clean energy transition. This method and results clearly show that a complete understanding of the channeling regime will be essential for any implementation of the advection-diffusion-reaction equations across a broad range of applications including flow organisation during magma melt \cite{2001-Spiegelman,2018-jones}, diagenesis \cite{hosa2020modelling,hosa2020order}, and other geological processes \cite{manga2001using}, drug delivery systems \cite{2015-mcginty}, contaminant transport in underground reservoirs \cite{2020-hasan,2020-pak, 2011-dentz}, and virus spreading dynamics \cite{2017-lin}. 

\section{Materials and Methods}
All numerical simulations were performed using \href{https://github.com/GeoChemFoam}{GeoChemFoam} on Intel Xeon processors (24 cores). For each image, an unstructured mesh is created within the pore-space using OpenFOAM utility \textit{snappyHexMesh}. For each time-step, velocity and acid concentration fields are solved. Then the reaction rate and the velocity of the dissolving faces are calculated and the mesh is updated. Mesh quality is checked at the end of each time-step and if the skewness is too large, the domain is completely remeshed. Since GeoChemFoam uses steady-state formulations of flow and transport, it can be applied with very large time-steps ($CFL\approx1000$), allowing for large speed-ups in computation time. Details of the geometry creation, analysis, meshing, numerical method, and time stepping strategy are presented in \cite{2022a-Maes} and in the supplementary material. The original geometries and output files can be downloaded from our \href{https://zenodo.org/record/6993528}{Zenodo dataset archive}, the geometry creation scripts are on \href{https://github.com/hannahmenke/Channeling2022}{github} and an example input deck is on the \href{https://github.com/GeoChemFoam/GeoChemFoam/tree/main/Examples/}{GeoChemFoam wiki}.

\subsection{Governing equations}

Under isothermal conditions and in the absence of gravitational effects, fluid motion in the pore-space is governed by the incompressible Navier-Stokes equations,
\begin{equation}\label{Eq:cont}\nabla\cdot\mathbf{u} = 0,
\end{equation}
\begin{equation}
\frac{\partial \mathbf{u}}{\partial t}+ \nabla\cdot\left(\mathbf{u}\otimes\mathbf{u}\right)=-\nabla p +\nu\nabla^2\mathbf{u},\label{Equ:momentum}
\end{equation}
with the continuity condition at the fluid-solid interface $\Gamma$,
\begin{equation}\label{Equ:bcu}
\rho\left(\mathbf{u}-\mathbf{w}_s\right)\cdot \mathbf{n}_{s}=-\rho_s\mathbf{w}_s\cdot\mathbf{n}_s \hspace{0.5cm} \text{at $\Gamma$},
\end{equation}
where $\mathbf{u}$ (m/s) is the velocity, $p$ (m$^2$/s$^2$) is the kinematic pressure, $\nu$ (m$^2$/s) is the kinematic viscosity, $\rho$ (kg/m$^3$) is the fluid density, $\rho_s$ (kg/m$^3$) is the solid density, $\mathbf{n_s}$ is the normal vector to the fluid-solid interface pointing toward the solid phase, and $\mathbf{w_s}$ (m/s) is the velocity of the fluid-solid interface, which is controlled  by the surface reaction rate $R$ (kmol/m$^2$/s) such that
\begin{equation}\label{Eq:Ws}
\mathbf{w}_s=\frac{M_{ws}}{\rho_s}R\mathbf{n}_s,
\end{equation}
where $M_{ws}$ is the molecular weight of the solid. The concentration $c$ (kmol/m$^3$) of a species in the system satisfies an advection-diffusion equation
\begin{equation}\label{Eq:concentration}
\frac{\partial c}{\partial t}+ \nabla \cdot \left( c\mathbf{u} \right) = \nabla\cdot\left(D\nabla c\right),
\end{equation}
where  $D$ (m$^2$/s) is the diffusion coefficient. The chemical reaction occurs at the fluid-solid interface $\Gamma$, such that
\begin{equation}\label{Equ:bcc1}
\left(c\left(\mathbf{u}-\mathbf{w}_s\right)-D\nabla c \right)\cdot \mathbf{n}_{s}=\zeta R \hspace{0.5cm} \text{at $\Gamma$},
\end{equation}
where $\zeta$ is the stoichiometric coefficient of the species in the reaction. In this work, we assume that the surface reaction rate depends only on the concentration of one reactant species, following
\begin{equation}
 R=k_cc,
\end{equation}
where $k_c$ (m/s) is the reaction constant. At the inlet, the boundary conditions are constant flow rate $Q$ (m$^3$/s) and constant reactant concentration $c_i$ (kmol/m$^3$). To limit inlet boundary effect, the velocity is extrapolated from a zero gradient rather than taken as constant \citep{2016-OpenFOAM}. At the outlet, the boundary conditions are constant pressure $p_0$ (m$^2$/s) and a zero gradient for velocity and reactant concentration.

\subsection{Dimensionless analysis}
The flow, transport and reaction conditions are characterized by the Reynolds number
\begin{eqnarray}
Re=\frac{UL}{\nu},
\end{eqnarray}
which quantifies the relative importance of inertial to viscous forces, the P\'eclet number,
\begin{eqnarray}
Pe=\frac{UL}{D},
\end{eqnarray}
which quantifies the relative importance of advective and diffusive transport, and the Kinetic number,
\begin{eqnarray}
Ki=\frac{k_cL}{D},
\end{eqnarray}
which quantifies the relative importance of chemical reaction and diffusive transport. Here $U$ and $L$ are the reference pore-scale velocity and length. The Kinetic number characterized if the chemical reaction at the surface of solid grains is in the reaction-limited ($Ki<1$) or transport-limited ($Ki>1$) regime. The Damk\"ohler number $Da$, which is the ratio of Kinetic and  P\'eclet numbers, is also a relevant quantity. $Da$ quantifies the relative importance of reaction to advective transport globally, but not locally as the reactant can only be transported to the solid surface by diffusion (Equ. (\ref{Equ:bcu}) and (\ref{Equ:bcc1})). In this study, we assume that we are in the creeping flow regime ($Re<<1$) so that the dissolution regime is only dependent on $Pe$ and $Ki$. In addition, the reactant strength, defined as
\begin{eqnarray}
 \beta=\frac{c_{i}M_{ws}}{\zeta\rho_s},
\end{eqnarray}
characterised how many kg of solid are dissolved by a kg of reactant.
The pore-scale reference velocity is chosen as the average pore velocity, defined as
\begin{eqnarray}
U=\frac{U_D}{\phi},
\end{eqnarray}
where $\phi$ is the porosity of the domain and $U_D$ (m/s) is the Darcy velocity, defined as
\begin{eqnarray}
U_D=\frac{Q}{A},
\end{eqnarray}
where $A$ (m$^2$) is the cross-sectional area of the domain. The pore-scale reference length scale $L$ is defined as
\begin{equation}
    L=\sqrt{\frac{12K}{\phi}},
\end{equation}
where $K$ (m$^2$) is the permeability of the domain, and the parameter 12 is a constant defined so that the pore-scale length scale corresponds to the tube size for a capillary bundle of constant size. The permeability can be calculated as
\begin{equation}\label{Darcy}
    K=-\frac{\nu U_DL_D}{\Delta P},
\end{equation}
where $L_D$ is the length of the domain and $\Delta P$ is the pressure drop between inlet and outlet. The pressure is a constant at the outlet, but not at the inlet (constant flow rate boundary condition). Therefore, the pressure drop is defined as \citep{2014-Raeini}
\begin{equation}\label{pressDrop}
    \Delta P = -\frac{1}{Q}\frac{dW_P}{dt},
\end{equation}
where $W_P$ is the work done by the pressure force in the domain. Equ. \ref{Darcy} and \ref{pressDrop} together denote that, for an equivalent flow rate, a higher permeability corresponds to a lower energy dissipation in the domain.  The rate of energy dissipation $\frac{dW_P}{dt}$ can be calculated as
\begin{equation}
\frac{dW_P}{dt}=-\int_V{\nabla p\cdot udV}.
\end{equation}

\subsection{Quasi-static assumption}

Dissolution of a solid grain is typically orders of magnitude slower than reactant transport. This is characterised in our numerical model by $\beta Da<<1$ and $\beta Ki<<1$. For example, for dissolution of calcite ($M_{ws}=100$ kg/kmol, $\rho_s=2710$ kg/m$^3$) by an acid at pH=2 ($c_i=0.01$ kmol/m$^3$), the reactant strength $\beta$ is equal to $3.69\times10^{-4}$. Therefore, as long as $Pe<100$ and $Ki<100$, the displacement of the solid interface is slow compared to the transport of reactant in the domain, and  flow (Equ. (\ref{Equ:momentum})) and transport (Equ. \ref{Eq:concentration})) can be assumed to be in a quasi-static state
\begin{equation}
\nabla\cdot\left(\mathbf{u}\otimes\mathbf{u}\right)=-\nabla p +\nu\nabla^2\mathbf{u},\label{Eq:momentumQS}
\end{equation}
\begin{equation}\label{Eq:concentrationQS}
\nabla \cdot \left( c\mathbf{u} \right) = \nabla\cdot\left(D\nabla c\right).
\end{equation}
The quasi-static assumption allows the models to run with a large time-step controlled only by the velocity of the solid interface to save on computational time.

\backmatter

\bmhead{Supplementary material}
Supplementary materials can be found at the end of this manuscript. Supplementary data can be found at \href{https://zenodo.org/record/6993528}{Zenodo dataset archive}, the geometry creation scripts are on \href{https://github.com/hannahmenke/Channeling2022}{github} and an example input deck is on the \href{https://github.com/GeoChemFoam/GeoChemFoam/tree/main/Examples/}{GeoChemFoam wiki}.

\bmhead{Acknowledgments}
This work was supported by the UK EPSRC funded project on Direct Numerical Simulation for Additive Manufacturing in Porous Media (EP/P031307/1) and by Energi Simulation. The authors would like to give special thanks to Professor Marc Spiegelman for insightful comments. 

\bibliography{sn-bibliography}


\begin{appendices}

\section*{Appendix A: Model A and Model B Geometries}
\subsection*{Geometry Creation}

A uniform geometry was created with a uniform bead radius of 12 pixels placed on a diagonal grid with a spacing of 40 pixels and an offset of 20 pixels. A small random deviation of 2 pixels in the placement of the beads and 4 pixels in the radius of the beads was then introduced into this homogeneous model to allow for preferential flow paths to develop (Model A). Structural complexity was then increased by creating another model (Model B) using the same grid, spacing, and offset, but with random deviation of 6 pixels in bead radius and 12 pixels in bead placement. The model was set on a 1200 x 1200 pixel image which was then output at 10 times the resolution to preserve edges as a 12000 x 12000 pixel image. This image was then binned by 12 in each direction using ImageJ and padded by 2 on every side using Python to give the final model dimensions of 1004x1004 pixels. The resolution of the geometry was set to 3.5 $\mu$m per pixel, giving a domain size of 3cm$\times$3cm. 

\begin{figure}
\includegraphics[width=\textwidth]{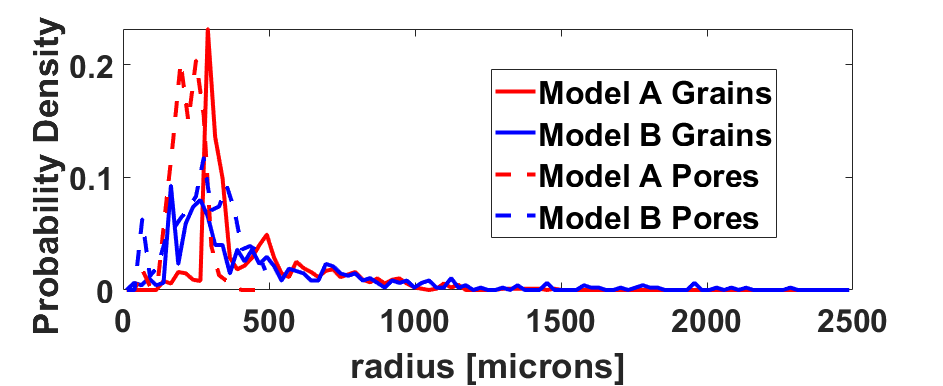}
\caption{The pore and grain radius distributions for the initial geometries of Model A and B.
\label{fig:PoreGrainPDF}}
\end{figure}

Each domain was meshed and the flow field calculated using the Open Source Computational Fluid Dynamics toolbox OpenFOAM \cite{2016-OpenFOAM} (Fig 2A \& B). The distribution of pore throat sizes and velocities are presented in Fig 2C and the distribution of pore and grains sizes are presented in Fig \ref{fig:PoreGrainPDF}. The scripts for creating the initial 12000x12000 geometries can be found on \href{https://github.com/hannahmenke/Channeling2022/}{github}. The original images with the radius, x, and y coordinates of each bead can be found on our \href{https://zenodo.org/record/6993528}{Zenodo dataset archive}.  

\subsection*{Geometry Analysis with Image Analysis}

\begin{figure}
\includegraphics[width=\textwidth]{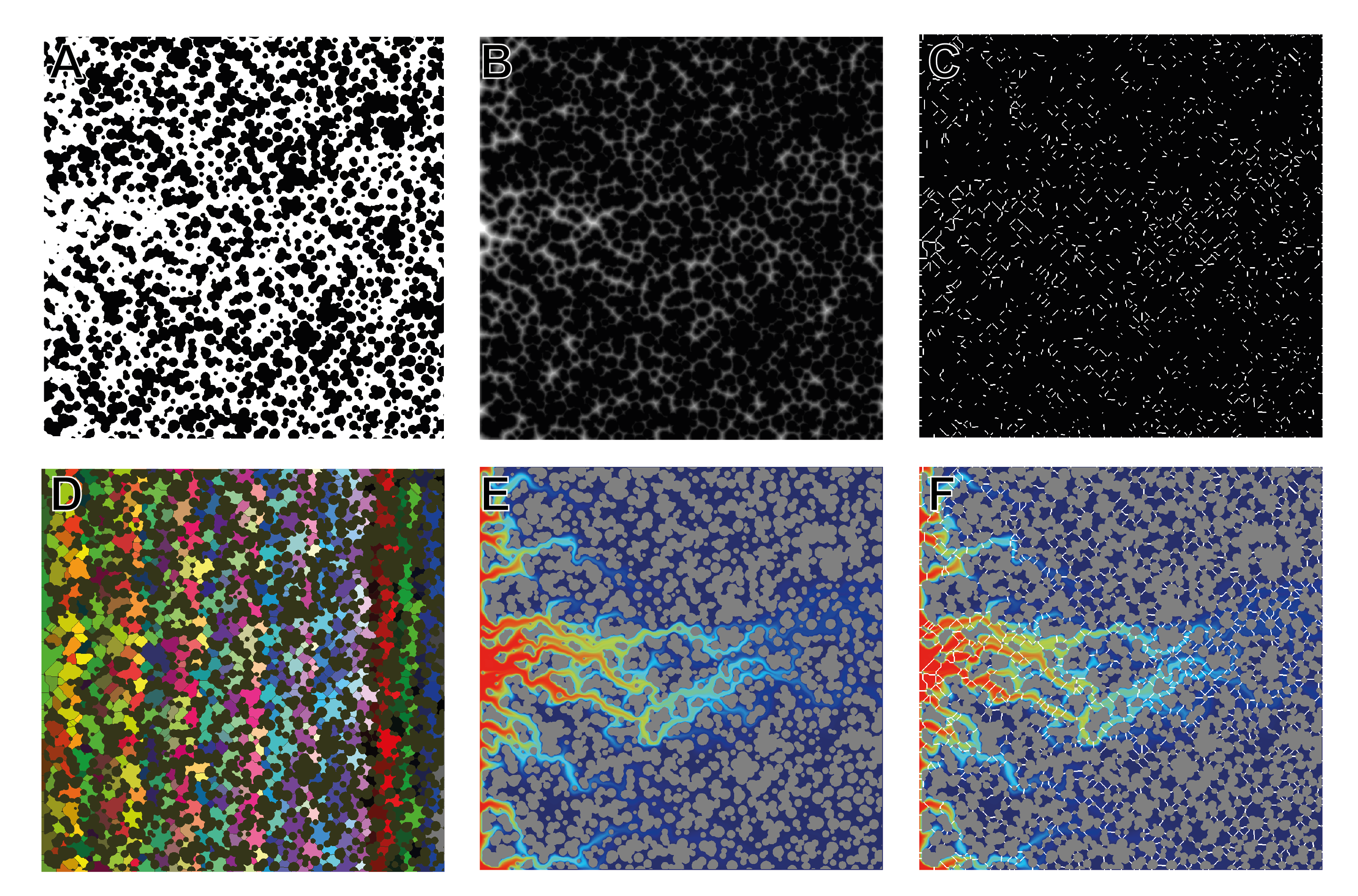}
\caption{(A) The pore space with pores in white and grains in black. (B) A Euclidean distance map was calculated on the pore space. (C) The local maxima of the distance map are designated the center of each pore. The boundaries between pores are designated as throats. (D) Each pore and throat is then individually identified, and local statistics calculated. (E) The concentration map (colored) is then overlain on the pore space with the grains in grey and (F) the local concentration statistics for each pore and throat are then calculated.
\label{fig:ImageProcessing}}
\end{figure}

The grains, pores, and pore throats were extracted from each time step in the simulations using a watershed segmentation algorithm and the Euclidean distance map of the grain and pore spaces was used to identify individual grains and pores with the boundaries between pores as throats. An example of this method with each initial geometry is shown in Fig \ref{fig:ImageProcessing}. The statistics of the grain, pore, and pore throat size distributions along with the characteristic length and velocities (at Pe=1) are given in Table \ref{Table:ModelProperties}. 

\begin{table}\centering
\caption{Table of initial geometry statistics\label{Table:ModelProperties}}

\begin{tabular}{lcc}
Statistic & Model A & Model B \\
\midrule
Characteristic Length $L$ [m] & 1.125 x $10^{-4}$ &  1.251 x x $10^{-4}$\\
Pore radius mean [pixels] & 6.4 & 8.3 \\
Pore radius standard deviation & 1.4 & 3.2 \\
Pore radius skewness & -0.23 & 0.19 \\
Pore radius kurtosis & 3.5 & 2.9 \\
Grain radius mean [pixels] & 12.9 & 14.3\\
Grain radius standard deviation  & 6.3 & 12.6 \\
Grain radius skewness & 1.5 & 2.6 \\
Grain radius kurtosis  & 5.4 & 12.4 \\
Pore throat radius mean [pixels] & 2.5 & 3.8 \\
Pore throat radius standard deviation & 1.2 & 2.4 \\
Pore throat radius skewness & 1.0 & 1.1\\
Pore throat radius kurtosis &  4.7 & 5.0 \\
Pore velocity $U$ mean  [m/s] & 8.9 x $10^{-6}$ & 8.0 x x $10^{-6}$\\
Pore velocity $U$ standard deviation  & 0.83 & 1.0 \\
Pore velocity $U$ skewness  & 2.6 & 3.5\\
Pore velocity $U$ kurtosis  & 12 & 20 \\
\bottomrule
\end{tabular}
\end{table}

\subsection*{Geometry Analysis with Autocorrelation}
\begin{figure}
\includegraphics[width=\textwidth]{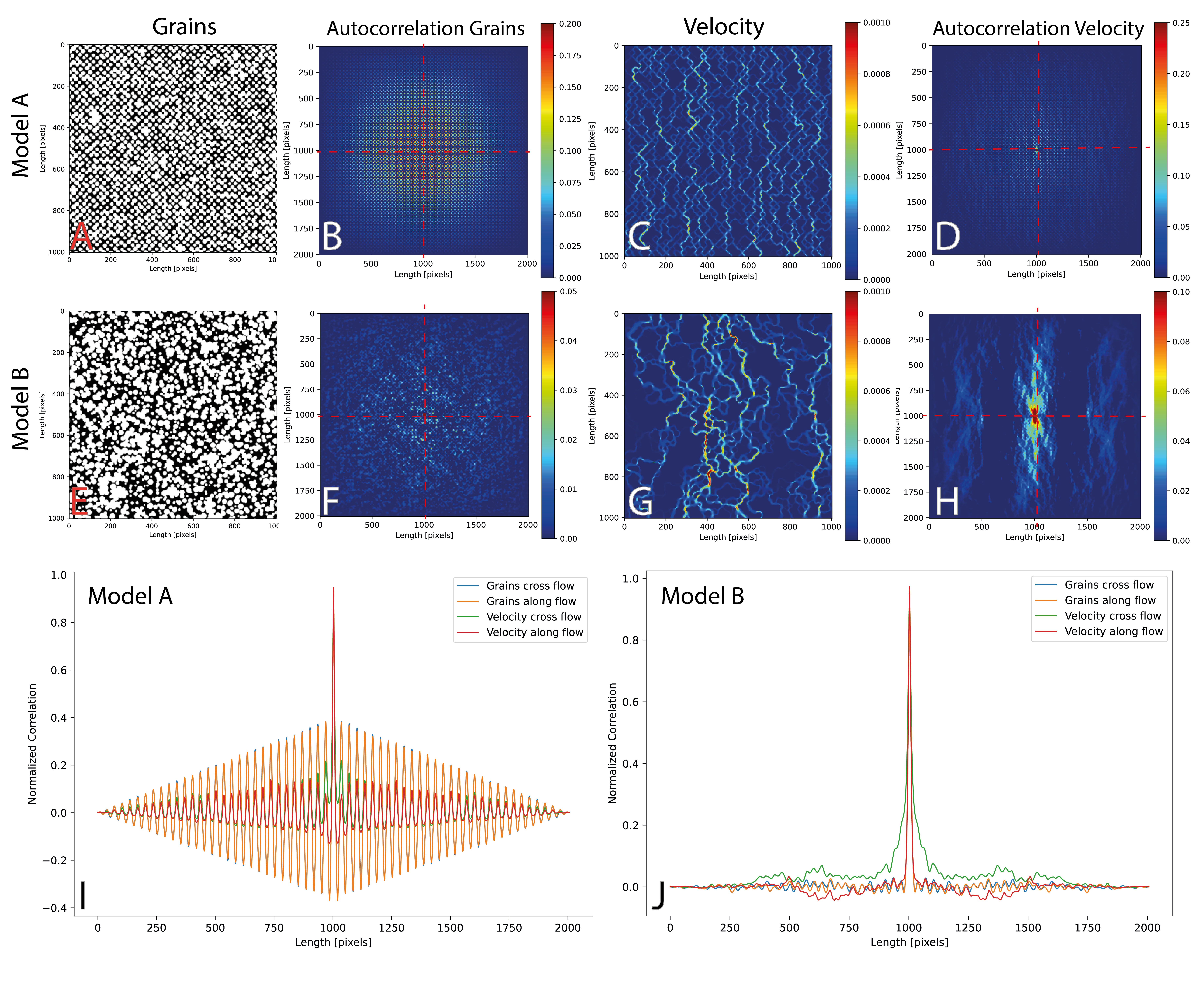}
\caption{The autocorrelation function for Model A and Model B shown for grains and velocity at Pe=1. A and E are the grains in white with the pores in black. B and F are the autocorrelation functions of the grains. C and G are the velocities in the direction of flow, D and H are the autocorrelation functions of the velocity. I and J are the autocorrelations of the grains and velocities for Model A and B respectively in each direction plotted from the centre points of the autocorrelations marked by red dotted lines on B, F, D, and H.
\label{fig:AutoCorrelation}}
\end{figure}

Here we compute the autocorrelation of the grains and velocities for both Model A and Model B (Fig \ref{fig:AutoCorrelation}). Both models have an autocorrelation function that steeply decreases towards zero with lag, over a length scale equal to the grain spacing.  Model A is statistically anisotropic, with an autocorrelation function with square symmetry and prominent sidelobes reflecting the underlying grid. Model B is statistically isotropic, with no sidelobes.

The autocorrelation function of the along-flow component of the velocity field is statistically anisotropic, with rectangular symmetry.  The scale length in the along-flow direction typically is similar to the grain spacing but is larger (by a factor of about five) in the cross-flow direction, as is expected for channels.  For Model A, the autocorrelation has sidelobes reflecting the underlying periodicity of the medium, with wavelength equal to the grain spacing.  The autocorrelation for Model B is similar, but without the sidelobes.

\newpage

\section*{Appendix B: Numerical method}

\subsection*{Meshing}

\begin{figure}[!t]
\begin{center}
\includegraphics[width=0.75\textwidth]{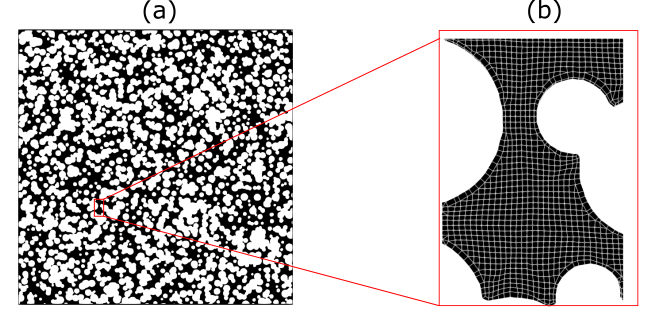}
\caption{Example of pore-space meshing (a) Full domain for model B (b) zoom and visualization of mesh inside a pore.\label{fig:mesh}}
\end{center}
\end{figure}

The equations are solved using finite volume discretization over an unstructured hybrid mesh. To build the mesh, the solid surface is described using an \textit{stl} image. First, a Cartesian mesh of resolution $h$ is generated. The mesh is then snapped onto the solid surface using the \textit{snappyHexMesh} utility \cite{2016-OpenFOAM}, i.e. cell containing solid are then removed and replaced by hexahedral or tetrahedral cells that match the solid boundaries. An additional layer of cells of the same resolution $h$ is then added around the solid boundary to improve the representation of the solid surface. To decide the resolution used for the initial mesh, a convergence study on porosity and permeability was conducted for Model B (Table \ref{Table:MeshConvergence}). We observe that a resolution of 3 $\mu$m offers a good compromise between accuracy and size of computational mesh. Fig. \ref{fig:mesh} shows Model B with a zoom into a pore to observe the mesh at resolution 3 $\mu$m. 

\begin{table}[!ht]
\centering
\caption{Mesh convergence (Model B)\label{Table:MeshConvergence}}
\begin{tabular}{lccc}
Resolution ($\mu$m) & Porosity & Permeabilty (m$^2$) & number of cells \\
\midrule
6 & 0.432 & 3.17$\times 10^{-10}$ & 139k\\
3 & 0.455 & 5.64$\times 10^{-10}$ & 526k\\
2 & 0.457 & 5.68$\times 10^{-10}$ & 1141k\\
\bottomrule
\end{tabular}
\end{table}

\subsection*{Arbitrary Lagrangian Eulerian method}

The equations are solved using the Arbitrary Lagrangian Eulerian (ALE) method \citep{2016-Starchenko}, implemented in GeoChemFoam (\href{www.github.com/geochemfoam}{www.github.com/geochemfoam}) and the full solution procedure is presented in Fig. \ref{fig:solutionProcedure}. For each time-step, the mesh points are moved with velocity $\mathbf{w}$, which satisfies the Laplace equations with boundary condition (Equ. (\ref{Eq:Ws}))
\begin{eqnarray}\label{Eq:w}
 \nabla\cdot D_m \nabla w_j = 0 \hspace{0.5cm} \text{j=x,y,z}\\
 w_j=\mathbf{w}_s\cdot \mathbf{e}_j \hspace{0.5cm} \text{at $\Gamma$},
\end{eqnarray}
where $D_m$ is the diffusivity of the mesh motion, $w_j$ is the j-directional component and $\mathbf{e}_j$ is the j-directional standard basis vector. With these equations, the mesh points will track the fluid-solid interface, and the mesh motion is diffused to avoid large volume ratio between neighbor cells. However, as the mesh points are displaced, the skewness of the mesh can increase and lead to failure of the transport solver. To avoid this, the mesh's skewness is checked at the end of each time-step, and the domain is fully remeshed upon failure.  After remeshing, the velocity, pressure and concentration fields are mapped to the new mesh. In addition, topological errors can appear when two faces of the same mineral grain overlap, leading to failure of the flow or transport solver. To avoid this, the faces which are fully located in a topological error are eliminated before remeshing. These collapsing faces are identified by the following condition: a face defined as faceI collapsed if a ray leading from its center following its normal vector pointing toward the solid phase meets another face defined as faceJ at a distance lower than the grid size, and faceI and faceJ do not intersect. Following this remeshing algorithm, our numerical simulations are stable and topological errors are eliminated.

\subsection*{Time-stepping strategy}

\begin{figure}[!b]
\centering
\includegraphics[width=0.8\textwidth]{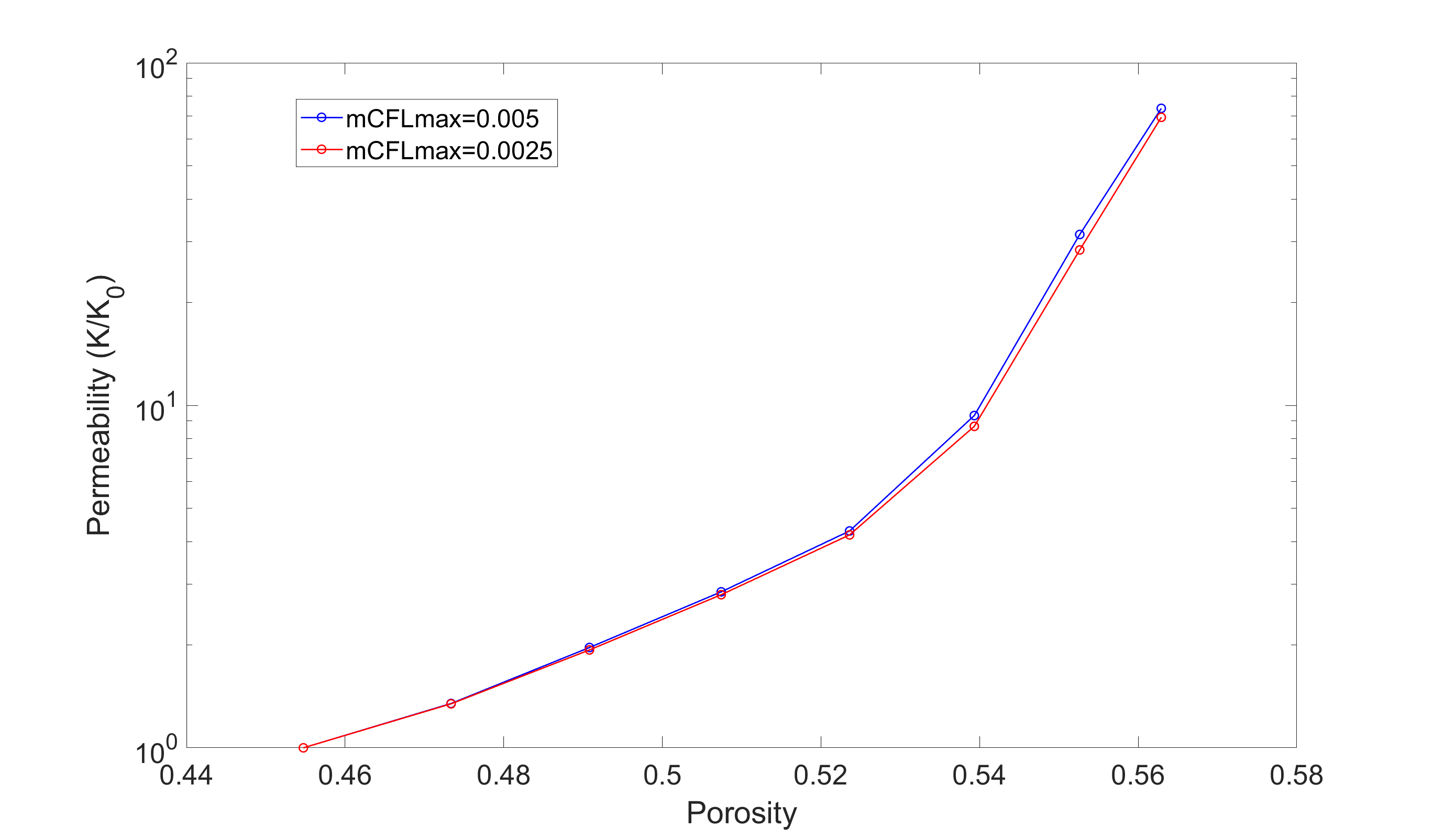}
\caption{Comparison of permeability evolution as a function of porosity for Model B at $Pe=1$, $Ki=1$ for two different maximum mCFL numbers. \label{fig:TimeStepConvergence}}
\end{figure}

\begin{figure}
\centering
\includegraphics[width=0.5\textwidth]{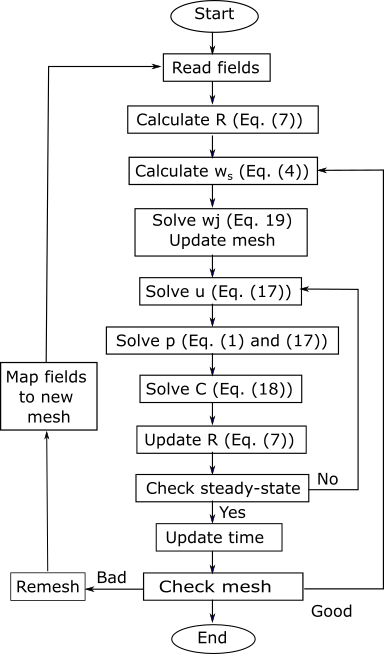}
\caption{Solution procedure for solving quasi-steady state dissolution using the ALE method. \label{fig:solutionProcedure}}
\end{figure}

The simulations are performed using an adaptive time-stepping strategy based on the mesh Courant-Friedrich-Lewy (CFL) number defined as
\begin{equation}
 mCFL = \frac{\mathbf{w}\Delta t}{h},
\end{equation}
where $\Delta t$ is the time-step and $h$ is the mesh resolution. The simulation are performed using a maximum $mCFL$ number of 0.005, which offers a good compromise between accuracy, robustness and efficiency. Fig. \ref{fig:TimeStepConvergence} shows a comparison of permeability evolution as a function of porosity for Model B at $Pe=1$, $Ki=1$ between mCFL=0.005 and mCFL=0.0025.

\section*{Appendix C: Robustness of $\phi$, $K$ and $L$ for stochastically generated micromodel}

The study presented in the paper is limited to one instance of each of two stochastic models (Model A and B). Future work will focus on extending the findings to any generated geometry and in particular on linking the dissolution regimes to the parameters of the stochastic distribution. For this, it would be essential that the geometrical parameters that are used in the calculation of $Pe$ and $Ki$, i.e. the porosity $\phi$, and the pore-scale length $L$, vary over a range much less than an order of magnitude, so that the calculation of $Pe$ and $ki$ are robust over different instance of the same stochastic distribution. Table \ref{Table:lengthscale} shows the variation of porosity and pore-scale length for 12 instances of each stochastic distribution (Model A and B). For model A, $\phi$ varies between 0.430 and 0.445 and $L$ varies between 1.04 and 1.15 $\times 10^{-4}$ m; for model B, $\phi$ varies between 0.451 and 0.473 and $L$ varies between 1.14 and 1.41 $\times 10^{-4}$ m. This shows that the calculation of $Pe$ and $Ki$ will be robust, as $\phi$ and $L$ varies on a scale much smaller than an order of magnitude. 

\begin{table}[!ht]
\centering
\caption{Porosity and $L$ for 12 realizations of Model A and B)\label{Table:lengthscale}}
\begin{tabular}{l|cc|cc}
Instance & \multicolumn{2}{c|}{Model A} & \multicolumn{2}{c}{Model B}  \\
\midrule
 & $\phi$ & $L$ ($\times10^{-4}$m) & $\phi$ & $L$ ($\times10^{-4}$m) \\
 1 & 0.437 & 1.11 & 0.455 & 1.22 \\
2  &  0.439  & 1.11 & 0.473 & 1.41  \\
3  & 0.436 & 1.11 & 0.455 & 1.34  \\
4 & 0.432 & 1.10 & 0.457 &	1.29  \\
5 & 0.445 & 1.13 & 0.466 &	1.28  \\
6 & 0.440 & 1.11 & 0.465 & 1.23  \\
7 & 0.436 & 1.11 & 0.464 & 1.31  \\
8 & 0.437 & 1.12 & 0.468 & 1.34  \\
9 & 0.439 & 1.15  & 0.460 &	1.31 \\
10 & 0.436 & 1.10 & 0.451 & 1.14  \\
11 & 0.430 & 1.04  & 0.463 & 1.25 \\
12 & 0.438 & 1.08 & 0.472 & 1.22  \\
\bottomrule
\end{tabular}
\end{table}

\section*{Appendix D: Time Sequence Videos of Dissolution}

Movies S1-8 show the dissolution time series for select simulations A1-A4 and B1-B4. 

\href{https://youtube.com/shorts/x-J1-x83y0E}{Movie S1}: Visualisation of Model A $Pe$=0.01 $Ki$=0.1 evolution of porosity and concentration. The grains are gray, with the concentration field in color. The pore throats are extracted by a watershed algorithm on the Euclidean distance map of the pore space and superimposed in white. This is an example of the compact dissolution regime. 

\href{https://youtube.com/shorts/gTIHaQsBaRA}{Movie S2}: Visualisation of Model A $Pe$=1 $Ki$=1 evolution of porosity and concentration. The grains are gray, with the concentration field in color. The pore throats are extracted by a watershed algorithm on the Euclidean distance map of the pore space and superimposed in white. This is an example of the wormhole formation dissolution regime.

\href{https://www.youtube.com/shorts/tQOhGYgEOWE}{Movie S3}: Visualisation of Model A $Pe$=100 $Ki$=10 evolution of porosity and concentration. The grains are gray, with the concentration field in color. The pore throats are extracted by a watershed algorithm on the Euclidean distance map of the pore space and superimposed in white. This is an example of the channeling dissolution regime.

\href{https://youtube.com/shorts/XMwYa4NCiaw}{Movie S4}: Visualisation of Model A $Pe$=100 $Ki$=0.1 evolution of porosity and concentration. The grains are gray, with the concentration field in color. The pore throats are extracted by a watershed algorithm on the Euclidean distance map of the pore space and superimposed in white. This is an example of the uniform dissolution regime.

\href{https://youtube.com/shorts/hVkntEKNz2U}{Movie S5}: Visualisation of Model B $Pe$=0.01 $Ki$=0.1 evolution of porosity and concentration. The grains are gray, with the concentration field in color. The pore throats are extracted by a watershed algorithm on the Euclidean distance map of the pore space and superimposed in white. This is an example of the compact dissolution regime.

\href{https://youtube.com/shorts/FNGScitflic}{Movie S6}: Visualisation of Model B $Pe$=1 $Ki$=1 evolution of porosity and concentration. The grains are gray, with the concentration field in color. The pore throats are extracted by a watershed algorithm on the Euclidean distance map of the pore space and superimposed in white. This is an example of the wormhole formation dissolution regime.

\href{https://www.youtube.com/shorts/CJKEcgAfN_c}{Movie S7}: Visualisation of Model B $Pe$=100 $Ki$=10 evolution of porosity and concentration. The grains are gray, with the concentration field in color. The pore throats are extracted by a watershed algorithm on the Euclidean distance map of the pore space and superimposed in white. This is an example of the channeling dissolution regime.

\href{https://youtube.com/shorts/Fwr5fxuAkZY}{Movie S8}: Visualisation of Model B $Pe$=100 $Ki$=0.1 evolution of porosity and concentration. The grains are gray, with the concentration field in color. The pore throats are extracted by a watershed algorithm on the Euclidean distance map of the pore space and superimposed in white. This is an example of the uniform dissolution regime.

\end{appendices}

\end{document}